\documentclass[12pt,onecolumn]{IEEEtran}
\usepackage{graphicx,epsfig,graphics,epsf}
\usepackage{float,amsmath,amsfonts,amssymb}

\def\CC{\mathbb C}

\def\EE{\mathbb E}
\def\IC{\mathcal I}
\def\PP{\mathbb P}

\def\NC{\mathcal N}
\def\UC{\mathcal U}

\def\Tr{\mathrm{Tr}}

\newtheorem{rem}{Remark}
\newtheorem{lem}{Lemma}

\begin{document}

\title{Diversity-Multiplexing Tradeoff\\ for the MIMO Static Half-Duplex Relay}
\author{Olivier L\'ev\^eque$^*$\footnote{$^*$ Olivier L\'ev\^eque is with the Facult\'e I\&C, Ecole Polytechnique F\'ed\'erale de Lausanne, 1015 Lausanne, Switzerland, olivier.leveque@epfl.ch}, Christophe Vignat$^{**}$\footnote{$^{**}$ Christophe Vignat is with the Institut Gaspar Monge, Universit\'e de Marne-la-Vall\'ee, 77454 Marne-la-Vall\'ee, France, vignat@univ-mlv.fr}, Melda Y\"uksel$^\ddagger$\footnote{$^\ddagger$ Melda Y\"uksel is with the EEE Department, TOBB University of Economics and Technology, 06560 Ankara, Turkey, yuksel@etu.edu.tr}
}

\maketitle

\begin{abstract}
In this work, we investigate the diversity-multiplexing tradeoff (DMT) of the multiple-antenna (MIMO) static half-duplex relay channel. A general expression is derived for the DMT upper bound, which can be achieved by a compress-and-forward protocol at the relay, under certain assumptions. The DMT expression is given as the solution of a minimization problem in general, and an explicit expression is found when the relay channel is symmetric in terms of number of antennas, i.e.~the source and the destination have $n$ antennas each, and the relay has $m$ antennas. It is observed that the static half-duplex DMT matches the full-duplex DMT when the relay has a single antenna, and is strictly below the full-duplex DMT when the relay has multiple antennas. Besides, the derivation of the upper bound involves a new asymptotic study of spherical integrals (that is, integrals with respect to the Haar measure on the unitary group $\UC(n)$), which is a topic of mathematical interest in itself.
\end{abstract}


\section{Introduction}
Next generation wireless communication will take place over large networks with multiple-antenna (MIMO) terminals. To be able to understand how to operate these complex networks optimally, we need a solid theoretical background. The diversity-multiplexing tradeoff (DMT) \cite{ZhengT03} provides such a theoretical background for multiple-antenna systems. It establishes the fundamental tradeoff between reliability and rate via diversity and multiplexing gains. Diversity gain is a measure of reliability, and shows how fast the error probability decays with increasing signal-to-noise ratio (SNR). Similarly, multiplexing gain is related to the transmission rate of the system, and shows how this rate increases with increasing SNR.

Cooperation/relaying is going to be fully integrated in the standard operation of next generation wireless communications~\cite{RelayTaskGroup}. In wireless channels, nearby nodes can overhear source messages for free. Due to this \textit{wireless broadcast advantage}, relays can process the overheard information and forward it to the destination terminal. The destination can then combine the direct signal from the source and the forwarded signals from the relays to improve the system performance \cite{CoverElG79}, \cite{SendonarisEA03_12}, \cite{LanemanTW02}.

To understand large, multiple-antenna cooperative networks, the theory has to account for practical system constraints such as power, bandwidth or delay limitations. Another important constraint in relay channels is the half-duplex constraint. As the transmit power overwhelms the received power, wireless devices cannot transmit and receive at the same time in the same band. In other words, they are not full-duplex; they have to operate in half-duplex mode.

The half-duplex relay channel DMT has been already studied in \cite{LanemanTW02}, \cite{AzarianEGS05}, \cite{PrasadV04}, \cite{BletsasKRL06}, \cite{PAT08} and \cite{YukselE06_J}. In \cite{LanemanTW02}, orthogonal amplify-and-forward and decode-and-forward DMT are studied for the single-antenna relay channel. In \cite{AzarianEGS05}, the authors study the single-antenna relay channel and investigate the DMT of the non-orthogonal amplify-and-forward and the dynamic decode-and-forward protocols. Prasad and Varanasi \cite{PrasadV04} investigate a multiple relay channel with single antenna source and relays and a multiple antenna destination and propose space-time coding strategies. Bletsas et al.~\cite{BletsasKRL06} study relay selection as an alternate way to obtain DMT improvements in a multiple relay setting, and Pawar et al.~describe in \cite{PAT08} a new ``quantize-and-map'' relaying scheme that achieves the half-duplex DMT. In all these works, the number of degrees of freedom in the source-destination channel is equal to one; i.e.~either the source or the destination is allowed multiple antennas, but not both. In \cite{YukselE06_J}, the authors study the multiple-antenna relay channel DMT, thereby increasing the degrees of freedom in the direct link. They first state the static half-duplex DMT upper bound without explicitly computing it, and then show that the compress-and-forward protocol achieves this upper bound for any number of antennas at the source, the relay or the destination.

The above mentioned papers enlighten many important issues about the DMT of the half-duplex relay channel. However, while compress-and-forward is known to be DMT achieving (under suitable assumptions on the available channel state information), the exact static half-duplex DMT for arbitrary antenna configurations has not been computed yet. When nodes have multiple antennas, this computation is a mathematically challenging problem at first sight, as it requires the knowledge of the joint eigenvalue distribution of two correlated Wishart matrices, which has not yet been addressed in the literature. In this paper, we take a slight detour that allows us to express the static half-duplex DMT as the solution of a minimization problem. An explicit expression is moreover found for the DMT when the source and the destination have $n$ antennas each and the relay has $m$ antennas. This analysis brings us to the study of new asymptotics of spherical integrals (see Section \ref{sec:spherical}), which is a topic of mathematical interest in itself.

Finally, a natural question regarding the half-duplex constraint on the relay is to understand whether this constraint imposes a limitation on the system performance, in comparison to a relay operating in full-duplex mode. The DMT computation leads us to the following rather unexpected answer: when the relay has two or more antennas, the half-duplex constraint imposes a strict limitation on the system performance, {\em irrespective} of the number of antennas at the source and the destination. The only case where the half-duplex constraint imposes no limitation (in the DMT sense) is therefore the case where the relay has a single antenna. This fact was already acknowledged in the case where source and destination have a single antenna each \cite{YukselE06_J}; it is nevertheless surprising to observe that when the relay has multiple antennas, increasing the number of antennas at both the source and the destination does not allow to get rid of the half-duplex limitation.

The paper is structured as follows. After the introduction of the system model in Section \ref{sec:systemmodel}, we first review in Section \ref{sec:prelim} classical DMT results for the point-to-point MIMO channel and the full-duplex relay; we then derive the cut-set upper bound on the DMT in the half-duplex case and describe the DMT achieving compress-and-forward scheme. In Section \ref{sec:method}, we present our computation method for the DMT upper bound, that involves the study of the joint eigenvalue distribution of random matrices and leads us to writing the DMT as the solution of a minimization problem. In Section \ref{sec:expression}, we give the explicit expression of the DMT in the case where the source and the detination have the same number of antennas. Finally, we analyze in detail in Section \ref{sec:spherical} the asymptotics of spherical integrals that are required for the DMT computation.


\section{System Model}\label{sec:systemmodel}
\begin{figure}[t]
\centering
\includegraphics[width=10cm]{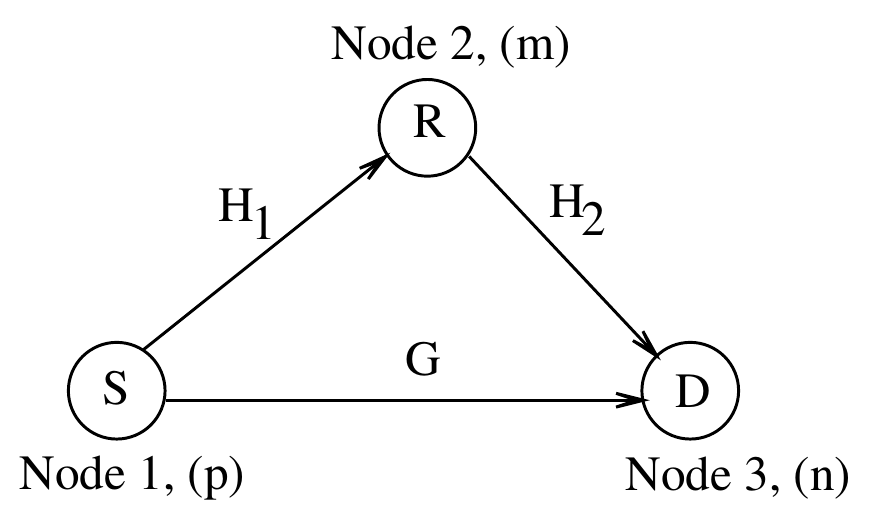} \caption{The half-duplex relay channel. The source, destination and relay have $p$, $m$ and $n$ antennas respectively. The relay listens for $t$ fraction of the time, and transmits in the remaining $1-t$.} \label{fig:systemmodelGH}
\end{figure}

We study the MIMO static half-duplex relay channel where the source, the relay and and the destination have $p$, $m$ and $n$ antennas respectively, see Figure \ref{fig:systemmodelGH}. The relay listens for a fraction $t$ of the time, and transmits in the remaining $(1-t)$ fraction, $t \in [0,1]$.

When the relay is listening, i.e.~the system is in state $q_1$, the received signals at the relay and the destination are
\begin{eqnarray*}
Y_{2,1} & = & H_{1} X_{1,1} + Z_{2,1} \\
Y_{3,1} & = & G X_{1,1} + Z_{3,1}
\end{eqnarray*}
When the relay is transmitting, i.e.~the system is in
state $q_2$, the received signal at the destination is
\begin{eqnarray*}
Y_{3,2} & = & G X_{1,2} + H_2 X_{2,2}+ Z_{3,2}
\end{eqnarray*}
The column vectors $X_{1,l}$ and $X_{2,l}$, $l = 1,2$ are respectively of length $n$ and $m$ and denote the signals the source and the relay transmit in state $q_l$. As the relay is half-duplex, $X_{2,1} = 0$. Similarly $Y_{2,l}$ and $Y_{3,l}$, $l = 1,2$ are respectively of length $m$ and $n$ and denote the received signals at the destination and at the relay. Note that $Y_{2,2} = 0$. The channel gain matrices $G$, $H_1$, and $H_2$ are of size $n \times p$, $m \times p$ and $n \times m$ respectively and are assumed to have independent and identically distributed (i.i.d.) zero mean complex Gaussian entries with unit variance. Fading is supposed slow and frequency non-selective, so that it remains fixed for one frame length. The relay and destination noise vectors $Z_{2,1}$ and $Z_{3,l}$ are of length $m$ and $n$ with i.i.d.~complex Gaussian entries with zero mean and unit variance.

\begin{rem}
If one considers the probably more realistic model where the variance of the entries of $G$, $H_1$ and $H_2$ are different for each matrix, the final result below concerning the diversity-multiplexing tradeoff does not change.
\end{rem}

The source and the relay have average short-term power constraints $P_1$ and $P_2$ for each codeword transmitted. As a constant scaling in transmitted power does not change the DMT \cite{ZhengT03}, we will assume that $P_1 = P_2 = P/2$
in the rest of this paper.

There is no transmitter channel state information at the source or at the relay. On the other hand, we assume that there are pilot signals to measure receiver channel gains, and thus the relay and the destination know their incoming fading levels.

In this paper, we compute the static half-duplex DMT upper bound, when there is only receiver channel state information. On the other hand, the bound presented in the following is still valid, if transmitter channel state information is available at the source, and the relay and the destination know all the channel gains in the system, but the system operates at \emph{constant information rate}. Under these assumptions, there exists a scheme that achieves the DMT upper bound we find. We describe this DMT optimal relaying protocol in Section \ref{subsec:PrelimAch}.

Although the relay is informed about its incoming channel gains, we assume that the relay does not use this information to determine the amount of time it listens or transmits. We therefore assume that the relay employs a {\emph{static protocol}} for communication, in which the decision of transmitting or receiving is not based on the channel matrix realizations, but only on their respective distributions. {\em Dynamic protocols}, in which the relay chooses $t$ based on its channel state observations, have the potential to achieve better half-duplex DMT. However, dynamic protocols are more complex to study and the half-duplex DMT computation already constitutes a hard problem. In this paper, we thus focus on static protocols only.

Without loss of generality, we also assume that the relay employs a fixed protocol. In fixed protocols, the relay does not carry information via breaking its transmission and reception intervals into smaller blocks and controlling
its state variable. If the relay protocol were not fixed but random, the state variable could be used to convey additional information. However, the increase is at most one bit and the DMT results remain the same \cite{YukselE06_J}.

As the source node does not have channel state information, an outage occurs if the mutual information at the destination is not large enough to support the fixed target communication rate the source chooses. In this paper, we study the minimum outage probability achieved by such a system at high SNR, following the approach of Zheng and Tse in \cite{ZhengT03}. In the next section, we overview the DMT for point-to-point MIMO channels, the DMT for the MIMO full-duplex relay channel, and the cut-set bound mutual information expressions for the MIMO static half-duplex relay channel.


\section{Preliminaries}\label{sec:prelim}

\subsection{Diversity-Multiplexing Tradeoff for the Point-to-Point MIMO Channel and the Full-Duplex Relay}
The DMT was first defined for slow, frequency non-selective Rayleigh fading point-to-point MIMO channels \cite{ZhengT03}. Assume there are $p$ antennas at the source and $n$ antennas at the destination. Let $R(P)$ denote the transmission rate of this $n \times p$ system and $\PP_e(P)$ denote the probability of error as a function of the transmit power (equal to the average received power as the channel gains have unit variance). We let the transmission rate vary logarithmically with the transmit power constraint. Then the multiplexing gain $r$ and the corresponding diversity gain $d$ are defined as
$$
\lim_{P \to \infty}\frac{ R(P)}{\log P} =r \quad \text{and} \quad
\lim_{P \to \infty}\frac{ \log \PP_e(P)}{\log P} = -d
$$
The DMT curve $d_{p,n}(r)$ indicates the tradeoff between these two gains, and is the piecewise linear curve joining the points $(k,d_{p,n}(k))$ for $0 \leq k \leq p \wedge n \triangleq \min\{p,n\}$, where
$$
d_{p,n}(k)=(p-k)(n-k), \quad k \in \{0,\ldots,p \wedge n \}
$$
Notice that $d_{p,n}(r) = d_{n,p}(r)$. For ease of notation, we will assume in the following that the transmission rate is chosen as $R(P)=r \log P$, for all $P$.

The full-duplex relay channel DMT was computed in \cite{YukselE06_J}. The relay forms a virtual antenna array with either the transmitter or the receiver, so using the classical cut-set bound argument shows that the performance of the system is upper bounded by the minimum of the performance of a MIMO point-to-point channel with either $(p+m) \times n$ antennas or $p \times (m+n)$ antennas, i.e.,
$$
d_{\mathrm{FD}}(r) = \min(d_{p+m,n}(r),d_{p,n+m}(r))
$$
Moreover, the compress-and-forward protocol achieves this DMT upper bound, when the relay and the destination are informed about all the channel gains in the system \cite{YukselE06_J}. In the following, we will use this performance as a benchmark and exhibit what loss is to be expected when operating the relay in half-duplex mode.


\subsection{Achievable Scheme for the Half-Duplex Relay}\label{subsec:PrelimAch}
In this paper, our objective is to compute the static half-duplex MIMO relay channel DMT under the assumptions explained in Section~\ref{sec:systemmodel}. However, no relaying protocol is known to achieve this bound under these assumptions. Only under full channel state information assumption at the relay and the destination do we know how to achieve the static half-duplex DMT upper bound. In such systems, the only known static half-duplex DMT optimal protocol is compress-and-forward.

In the compress-and-forward protocol, the relay first compresses its received signal $Y_{2,1}$, and then forwards the compressed signal through the relay-destination channel in the $1-t$ fraction of the time it transmits. The compression at the relay is of Wyner- Ziv type \cite{WynerZ76}, in the sense that the relay compresses its received signal taking into account that the destination has side information $Y_{3,1}$, available directly from the source. Although this protocol does not utilize the channel state information available at the source, the relay needs to know all the channel gains in the system to ensure that the compressed signal is received reliably at the destination. The soft information transmission of the compress-and-forward protocol is crucial to achieve optimal diversity and multiplexing gains. Protocols such as decode-and-forward \cite{CoverElG79,KhojastepourSA03_2}, which perform a hard decision about the source message, result in DMT suboptimal performance.

As a side remark, we notice that the compress-and-forward operation is DMT optimal not only for the static half-duplex relay channel, but also for the dynamic one. However, for the dynamic half-duplex relay channel, the DMT upper bound is hard to compute.



\subsection{Cut-Set Upper Bound for the Half-Duplex Relay}
In \cite{YukselE06_J}, it is shown that cut-set bounds and the associated outage expressions can be used to find DMT upper bounds. Following the same approach, we provide next the cut-set bounds for the static half-duplex relay channel and the corresponding probability of outage expressions to find the best half-duplex relay channel DMT for static protocols.

In the half-duplex, fixed and static relay channel, the mutual information expressions for the cut-sets around the source and the destination are respectively equal to
\begin{eqnarray}
I_S(t)& = & t I(X_1;Y_2 Y_3|q_1) +
(1-t)I(X_1;Y_3|X_2,q_2)
\label{eqn:cutset1}\\
I_D(t) & = & t I(X_1;Y_3|q_1) + (1-t)I(X_1 X_2; Y_3|q_2)
\label{eqn:cutset2}
\end{eqnarray}
To find the best static DMT upper bound, we need to find the maximum of these mutual
information expressions, which are obtained when Gaussian codebooks are used and the input covariance matrix is chosen optimally \cite{KhojastepourSA03_2,Telatar99}.

We know that for any channel matrix $H$ of size $n \times m$ and for any input covariance matrix $Q$ of size $m \times m$ \cite{ZhengT03},
\begin{equation}
\sup_{Q \geq 0, \mathrm{Tr}\{Q\}\leq {P}} \log \det
\left( I_n + H Q H^*\right) \leq
\log \det \left( I_n + P H H^* \right) \label{eqn:covarianceremoval}
\end{equation}
where $I_n$ denotes the identity matrix of size $n \times n$, and $^*$ denotes conjugate transpose. Using this inequality, we can further upper bound (\ref{eqn:cutset1}) and (\ref{eqn:cutset2}) as $I_S(t) \leq I_S'(t)$ and $I_D(t) \leq I_D'(t)$, where
\begin{eqnarray}
I_S'(t) & = & t \log \det \left( I_{m+n} + P \left[\begin{array}{c} G \\ H_1
\end{array}\right]\left[\begin{array}{c} G \\ H_1
\end{array}\right]^*\right) {+}\: (1-t) \log \det\left( I_n + PGG^*\right) \label{eqn:mutinfoS}\\ & & \nonumber \\
I_D'(t) &=& t \log \det\left( I_n + PGG^*\right) {+}\: (1-t) \log \det \left( I_n + P [G, H_2] \, [G,H_2]^* \right)\label{eqn:mutinfoD}
\end{eqnarray}
Thus, the outage probability corresponding to a target rate $r \log P$ is lower bounded by
\begin{equation}
\PP_{\mathrm{out}}(r \log P) \geq \min_{t \in [0,1]}  \max \left\{ \PP (I_S'(t)< r \log P), \PP (I_D'(t)< r \log P)\right\}
\triangleq \PP_{\mathrm{out},0}(r \log P) \label{eqn:PoutR0}
\end{equation}
These expressions lead to the DMT upper bound
\begin{equation}
d_{\mathrm{HD}}(r) \leq \max_{t \in [0,1]} \min \{ d_S(r,t),
d_D(r,t)\} \triangleq  d_{\mathrm{HD},0}(r) \label{eqn:DMT_upperbound}
\end{equation}
where $\PP_{\mathrm{out}}(r \log P) \dot{=} P^{-d_{\mathrm{HD}}(r)}$, $\PP(I_S'(t) < r \log P) \dot{=} P^{-d_S(r,t)}$, $\PP(I_D'(t) < r \log P) \dot{=} P^{-d_D(r,t)}$\\ and $\PP_{\mathrm{out},0}(r \log P) \allowbreak \dot{=} P^{-d_{\mathrm{HD},0}(r)}$~\footnote{Note that $f(P)\dot{=}P^{-c}$ means $\lim_{P \rightarrow \infty} \log f(P) / \log P = c$. Inequalities are defined similarly.}.

When the relay is half-duplex, it can only transmit during a fraction $t \in [0,1]$ of the time (and therefore receive during the other fraction $1-t$). Because of the static protocol assumption, the fraction $t$ is a fixed number, chosen according to the distribution of the channel coefficients only, and not to their realizations (but notice that $t$ depends a priori on the target rate $r$).

In Section \ref{sec:method} below, we explain how to write the above diversity order $d_{\mathrm{HD},0}(r)$ as the solution of a minimization problem, following the methodology of \cite{ZhengT03}. In Section \ref{sec:expression}, we show that in the particular case where the number of antennas at the source and the destination are equal (i.e., $p=n$), the symmetry of the above minimization problem allows us to write down $d_{\mathrm{HD},0}(r)$ explicitly.


\section{Computation Method} \label{sec:method}
In this section, our aim is to compute the diversity orders corresponding to the following outage probabilities
\begin{eqnarray*}
\lefteqn{\PP (I_S'(t)< r \log P)}\\
& = & \PP\left( t \log \det \left( I_{m+n} + P \left[\begin{array}{c} G \\ H_1 \end{array}\right]
\left[\begin{array}{c} G \\ H_1 \end{array}\right]^*\right) {+}\: (1-t) \log \det\left( I_n + PGG^*\right) < r \log P \right)
\end{eqnarray*}
and 
$$
\PP (I_D'(t)< r \log P) = \PP \left( t \, \log \det\left( I_n + P GG^*\right) + (1-t) \, \log \det \left( I_n + P [G,H_2] \, [G,H_2]^* \right) < r \log P \right)
$$
for a given $t \in [0,1]$. As the dimensions $p$, $m$ and $n$ are arbitrary,
it is sufficient to consider the second case; the first one will be deduced
correspondingly. For ease of notation, let us also write $H_2=H$ in the following,
so
$$
\PP (I_D'(t)< r \log P) = \PP \left( t \, \log \det(I_n + PGG^*) + (1-t) \, \log \det(I_n + PGG^* + PHH^*) < r \log P \right)
$$
where $G$, $H$ are two independent matrices, each with i.i.d.~$\NC_\CC(0,1)$ entries, $G$ is $n \times p$ and $H$ is $n \times m$. Remember also that the diversity order is defined as
\begin{equation} \label{cutsetD}
d_D(r,t) = - \lim_{P \to \infty} \frac{\log (\PP (I_D'(t) < r \log P)}{\log P}, \quad r \in [0,n]
\end{equation}
Before entering into the detailed computation, let us make the following observation. Let $\lambda_1,\ldots,\lambda_n$ be the eigenvalues of $GG^*$ and $\nu_1,\ldots,\nu_n$ be the eigenvalues of $GG^*+HH^*$. Since both these matrices are classical Wishart matrices, the (separate) joint distributions of $(\lambda_1,\ldots,\lambda_n)$ and $(\nu_1,\ldots,\nu_n)$ are well known. Nevertheless, in order to compute the above diversity order (following the methodology of Zheng and Tse in \cite{ZhengT03}), what is a priori needed is the {\em joint} distribution of {\em all} the eigenvalues $(\lambda_1,\ldots,\lambda_n,\nu_1,\ldots,\nu_n)$. This distribution is hard to obtain, because of the intricate correlation of the matrices $GG^*$ and $GG^*+HH^*$. A method is presented below that allows to compute the diversity order in the general case, but avoids the computation of the joint eigenvalue distribution. The latter question has not been addressed yet in the mathematical literature and remains an interesting open problem in its own right.

Our computation method goes as follows. We have
\begin{eqnarray*}
\lefteqn{t \, \log \det(I_n+PGG^*) + (1-t) \, \log \det(I_n+PGG^*+PHH^*)}\\
& = & \log \det(I_n+PGG^*) + (1-t) \, \log \det(I_m+PH^*(I_n+PGG^*)^{-1}H)
\end{eqnarray*}
Since $G$ and $H$ are independent and their respective distributions are unitarily invariant, this expression has the same distribution as
$$
\log \det(I_n+P\Lambda) + (1-t) \, \log \det(I_m+PH^*(I_n+P\Lambda)^{-1}H)
$$
where $\Lambda=\text{diag}(\lambda_1,\ldots,\lambda_{n \wedge p},0,\ldots,0)$, $\lambda_1 \ge \ldots \ge \lambda_{n\wedge p}$ are the non-zero eigenvalues of $GG^*$ and the number of zeros in the list is equal to $(n-p)^+$. Let now $B = (I_n+P\Lambda)^{-1/2} H$ ($n \times m$ matrix) and observe that
$$
\log \det(I_m+PH^*(I_n+P\Lambda)^{-1}H) = \log \det(I_m + B^*B) = \log \det(I_n + BB^*)
$$
{\em Conditioned on $\Lambda$} (which is independent of $H$), the joint distribution of the entries of $B$ is given by
$$
p(B|\Lambda) = \frac{1}{\pi^{nm}} \, \det(I_n+P\Lambda)^m \, \exp(-\Tr(B^*(I_n+P\Lambda)B))
$$
The spectral decomposition of the $n \times n$ matrix $BB^*$ reads $BB^*=U M U^*$, where $U$ is a $n \times n$ unitary matrix and $M=\text{diag}(\mu_1,\ldots,\mu_{n \wedge m},0, \ldots, 0)$, where $\mu_1 \ge \ldots \ge \mu_{n \wedge m}$ are the non-zero eigenvalues of $BB^*$ and the number of zeros in the list is equal to $(n-m)^+$.

As the diagonal matrices $\Lambda$ and $M$ might contain zeros, let us define $\tilde{\Lambda}=\text{diag}(\lambda_1,\ldots,\lambda_{n \wedge p})$ as well as
$\tilde{M}=\text{diag}(\mu_1,\ldots,\mu_{n \wedge m})$. The Jacobian of the transformation
$B \mapsto (\tilde{M},U)$ is given by
$$
J(\tilde{M},U) = (\det \tilde{M})^{|m-n|} \, \Delta(\tilde{M})^2, \quad \text{where }
\Delta(\tilde{M}) = \prod_{j,k=1 \atop j<k}^{n \wedge m} (\mu_j-\mu_k)
$$
Therefore, conditioned on $\Lambda$ (or $\tilde{\Lambda}$), the joint distribution of $(\tilde{M},U)$ is given by
$$
p(\tilde{M},U|\tilde{\Lambda}) = C_{n,m} \, \det(I_{n \wedge p} + P \tilde{\Lambda})^m \, \exp(-\Tr(UMU^*(I_n+P\Lambda))) \, (\det \tilde{M})^{|m-n|} \, \Delta(\tilde{M})^2
$$
so $U$ is independent of $\tilde{M}$ and distributed according to the Haar measure on the set $\UC(n)$ of unitary $n \times n$ matrices (that is, the columns of $U$ form a set of $n$ orthonormal vectors which are uniformly distributed on the sphere $\{z \in \CC^n \, : \, |z|=1\}$).

This allows us to compute the conditional distribution $p(\tilde{M}|\tilde{\Lambda})$:
\begin{eqnarray*}
\lefteqn{p(\tilde{M}|\tilde{\Lambda}) = \int_{\UC(n)} dU \, p(\tilde{M},U|\tilde{\Lambda})}\\
& = & C_{n,m} \, \det(I_{n \wedge p} +P \tilde{\Lambda})^m \, (\det \tilde{M})^{|m-n|} \, \Delta(\tilde{M})^2 \int_{\UC(n)} dU \, \exp(-\Tr(UMU^*(I_n+P\Lambda)))\\
& = & C_{n,m} \, \det(I_{n \wedge p} +P \tilde{\Lambda})^m \, (\det \tilde{M})^{|m-n|} \, \Delta(\tilde{M})^2 \, \exp(-\Tr(\tilde{M})) \int_{\UC(n)} dU \, \exp(-P \, \Tr(UMU^*\Lambda))
\end{eqnarray*}
Let finally
$$
\Delta(\tilde{\Lambda})=\prod_{j,k=1 \atop j<k}^{n \wedge p} (\lambda_j-\lambda_k)
$$
It is a well known fact that the distribution of $\tilde{\Lambda}$ is the classical Wishart distribution:
$$
p(\tilde{\Lambda}) = C_{n,p} \, (\det \tilde{\Lambda})^{|n-p|} \, \Delta(\tilde{\Lambda})^2 \, \exp(-\Tr(\tilde{\Lambda}))
$$
so we obtain
\begin{eqnarray}
p(\tilde{\Lambda},\tilde{M}) & = & p(\tilde{\Lambda}) \, p(\tilde{M}|\tilde{\Lambda}) \; = \; C_{n,m,p} \, \det(I_{n \wedge p} +P \tilde{\Lambda})^m \, (\det \tilde{\Lambda})^{|n-p|} \, (\det \tilde{M})^{|m-n|}\nonumber\\
& & \times \Delta(\tilde{\Lambda})^2 \, \Delta(\tilde{M})^2 \, \exp(-\Tr(\tilde{\Lambda}+\tilde{M})) \, \int_{\UC(n)} dU \, \exp(-P \Tr(UMU^*\Lambda)) \label{joint}
\end{eqnarray}
Notice that the spherical integral on the far right may be rewritten as
$$
\int_{\UC(n)} dU \, \exp(-P \, \Tr(UMU^*\Lambda)) = \EE_U \left( \exp \left( - P \sum_{j=1}^{n \wedge p} \sum_{k=1}^{n\wedge m} \lambda_j \, \mu_k \, |u_{jk}|^2 \right) \right)
$$
where $\EE_U$ denotes the expectation with respect to the Haar measure on $\UC(n)$.

Remember now that we are interested in computing
\begin{eqnarray*}
\PP_{\mathrm{out},0}(r \log P) & = & \PP \left( t \, \log \det(I_n + PGG^*) + (1-t) \, \log \det(I_n + PGG^* + PHH^*)  < r \log P \right)\\
& = & \int_{\{\tilde{\Lambda},\tilde{M} \, : \, \log\det(I_{n \wedge p} + P \tilde{\Lambda}) + (1-t) \,  \log\det(I_{n \wedge m} + P \tilde{M}) < r \log P\}} p(\tilde{\Lambda},\tilde{M}) \, d\tilde{\Lambda} \, d\tilde{M}
\end{eqnarray*}
Following Zheng and Tse, let us make the change of variables $\lambda_j=P^{-\alpha_j}$ and $\mu_k=P^{-\beta_k}$, where $\alpha_1 \le \ldots \le \alpha_{n \wedge p}$ and $\beta_1 \le \ldots \le \beta_{n \wedge m}$. The behaviour in the limit $P \to \infty$ of most terms in (\ref{joint}) is well known, except for the spherical integral. Let us define
$$
\IC(P) = \EE_U \left( \exp \left( -\sum_{j=1}^{n \wedge p} \sum_{k=1}^{n \wedge m} P^{1-\alpha_j-\beta_k} \, |u_{jk}|^2 \right) \right)
$$
The asymptotic behaviour of $\IC(P)$ is given by
\begin{equation} \label{asymptotic1}
\lim_{P \to \infty} \frac{\log(\IC(P))}{\log P} = \left\{\begin{array}{l} -\infty, \quad \quad \text{if }
\exists j \le n \wedge p, \, k \le n \wedge m \text{ such that } j+k=n+1 \text{ and } \alpha_j+\beta_k<1\\ \\
\displaystyle - \sum_{j=1}^{n \wedge p} \sum_{k=1}^{n \wedge m} (1-\alpha_j-\beta_k)^+, \quad \text{otherwise}
\end{array} \right.
\end{equation}
We relegate the proof of this asymptotic equality to Section \ref{sec:spherical}.

The diversity order corresponding to the cut around the destination (\ref{cutsetD}) can now be computed via the standard method developed in \cite{ZhengT03}. The computation leads to the following optimization problem:
\begin{eqnarray}
d_D(r,t) & = & \min \sum_{j=1}^{n \wedge p} (n+p-2j+1) \, \alpha_j +  \sum_{k=1}^{n \wedge m} (n+m-2k+1) \, \beta_k \nonumber \\& & \quad \quad - m \sum_{j=1}^{n \wedge p} (1-\alpha_j)^+ + \sum_{j=1}^{n \wedge p} \sum_{k=1}^{n \wedge m} (1-\alpha_j-\beta_k)^+ \label{optprobD}
\end{eqnarray}
where the minimization takes place over the set of variables $\alpha_{n \wedge p} \ge \ldots \ge \alpha_1 \ge 0$ and $\beta_{n \wedge m} \ge \ldots \ge \beta_1 \ge 0$ such that
$$
\sum_{j=1}^{n \wedge p} (1-\alpha_j)^+ + (1-t) \, \sum_{k=1}^{n \wedge m} (1-\beta_k)^+ \le r
$$
and $\alpha_j + \beta_k \ge 1$, for all $j \in \{1,\ldots,n \wedge p\}$, $k \in \{1,\ldots,n \wedge m\}$ such that $j+k=n+1$.

Similarly, the diversity order corresponding to the cut around the source is given by
\begin{eqnarray}
d_S(r,t) & = & \min \sum_{j=1}^{p \wedge n} (p+n-2j+1) \, \alpha_j +  \sum_{k=1}^{p \wedge m} (p+m-2k+1) \, \beta_k \nonumber \\& & \quad \quad - m \sum_{j=1}^{p \wedge n} (1-\alpha_j)^+ + \sum_{j=1}^{p \wedge n} \sum_{k=1}^{p \wedge m} (1-\alpha_j-\beta_k)^+ \label{optprobS}
\end{eqnarray}
where the minimization takes place over the set of variables $\alpha_{p \wedge n} \ge \ldots \ge \alpha_1 \ge 0$ and $\beta_{p \wedge m} \ge \ldots \ge \beta_1 \ge 0$ such that
$$
\sum_{j=1}^{p \wedge n} (1-\alpha_j)^+ + t \, \sum_{k=1}^{p \wedge m} (1-\beta_k)^+ \le r
$$
and $\alpha_j + \beta_k \ge 1$, for all $j \in \{1,\ldots,p \wedge n\}$, $k \in \{1,\ldots,p \wedge m\}$ such that $j+k=p+1$.

In the next section, we solve the above minimizations problem in the particular case where $n=p$ and thus find an explicit expression for the diversity order $d_{\mathrm{HD},0}(r)$.


\section{Explicit Expression for the Diversity-Multiplexing Tradeoff in the Case of Equal Number of Antennas at the Source and the Destination} \label{sec:expression}

In the case where the number of antennas at the source and the destination are equal (i.e., $p=n$), the symmetry of the problem implies that the optimal value of (\ref{eqn:PoutR0}) and thus of (\ref{eqn:DMT_upperbound}) is reached when $t=1/2$, for all values of $r$, and the minimization problems (\ref{optprobD}) and (\ref{optprobS}) boil down to the same expression:
\begin{eqnarray}
d_D(r,1/2) = d_S(r,1/2) & = & \min \sum_{j=1}^n (2n-2j+1) \, \alpha_j +  \sum_{k=1}^{n \wedge m} (n+m-2k+1) \, \beta_k \nonumber \\& & \quad \quad - m \sum_{j=1}^n (1-\alpha_j)^+ + \sum_{j=1}^n \sum_{k=1}^{n \wedge m} (1-\alpha_j-\beta_k)^+ \label{optprob}
\end{eqnarray}
where  $\alpha_n \ge \ldots \ge \alpha_1 \ge 0$ and $\beta_{n \wedge m} \ge \ldots \ge \beta_1 \ge 0$ are such that
$$
\sum_{j=1}^n (1-\alpha_j)^+ + \frac{1}{2} \, \sum_{k=1}^{n \wedge m} (1-\beta_k)^+ \le r
$$
and $\alpha_j + \beta_k \ge 1$, for all $j \in \{1,\ldots,n\}$, $k \in \{1,\ldots,n \wedge m\}$ such that $j+k=n+1$. So
$$
d_{\mathrm{HD},0}(r) = d_S(r,1/2) = d_D(r, 1/2)
$$
The above minimization problem constitutes a convex programming problem. It therefore has a unique solution, which can be found via different classical algorithms. The solution given below has first been obtained through numerical simulations for a relatively large number of values of $n$ and $m$ ($1 \le n \le 5$ and $1 \le m \le 12$). From these results, the general form of the solution could then be deduced by extrapolation. We therefore acknowledge that our derivation is not completely analytical. Our confidence in the result is strong, however, and we check below that our proposed solution satisfies the constraints of the problem, so that it is at least an upper bound on the actual solution of the minimization problem.

In order to describe the solution, let us define $l_0$ to be the minimum of $n$ and $\lfloor \frac{m+1}{3} \rfloor$ (so $l_0=n$ if $m \ge 3n-1$; otherwise, $l_0$ is an integer number between $0$ and $n-1$). It turns out that this number delimitates three different regimes for the DMT curve:

a)  For $0 \le r \le l_0/2$ (low multiplexing gain), the outage probability of the half-duplex relay is determined by the outage of the $n \times m$ relay link matrix $H$ (the direct link matrix $G$ being already off). In this regime, the corner points of the diversity curve are given by
$$
d_{\mathrm{HD},0}(l/2) = n^2 + (m-l)(n-l), \quad \text{where }l \in \{0,\ldots,l_0\}
$$
b) For $l_0/2 \le r \le n-l_0/2$ (intermediate multiplexing gain; notice that this regime does not exist if $l_0=n$, that is, if $m \ge 3n-1$), the outage probability is determined by a combination of direct link $G$ and relay link $H$ outages, and
the corner points of the diversity curve are given by
$$
d_{\mathrm{HD},0}(l_0/2+l) = l_0^2 + (n+m-l)(n-l_0-l), \quad \text{where }l \in \{0,\ldots,n-l_0\}
$$
c) For $n-l_0/2 \le r \le n$ (high multiplexing gain), the outage probability is determined by the outage of the $n \times n$ direct link matrix $G$ only, and the corner points of the diversity curve are given by
$$
d_{\mathrm{HD},0}(n-l/2) = l^2, \quad \text{where }l \in \{0,\ldots,l_0\}
$$
Finally, the diversity curve $d_{\mathrm{HD},0}(r)$ is the (convex and) piecewise linear curve interpolating between all these corner points.

Notice that, as already mentioned in the introduction, the only case where the half-duplex curve matches the full-duplex $(m+n) \times n$ curve is when $l_0=0$, i.e. $m=1$, and $n$ takes any integer value. For any $m \ge 2$, the half-duplex curve does not match (and is therefore strictly below) the full-duplex curve. This can be readily checked by noticing that the horizontal positions of the corner points do not match: in the half-duplex case (for $m \ge 2$), some corner points are located at non-integer multiplexing gains.

\begin{figure}[t]
\centering
\includegraphics[width=4.5cm]{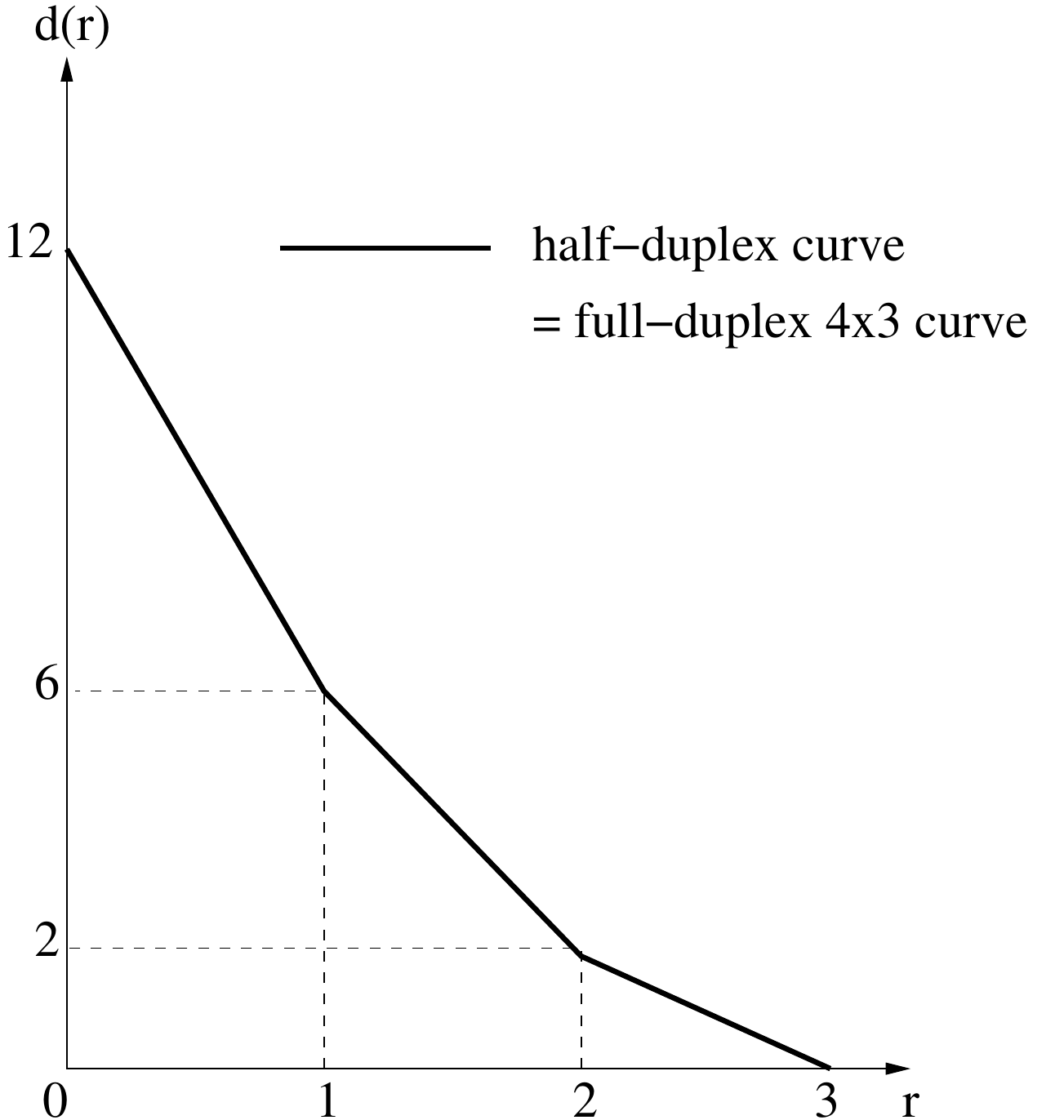} \hspace{1cm} \includegraphics[width=5cm]{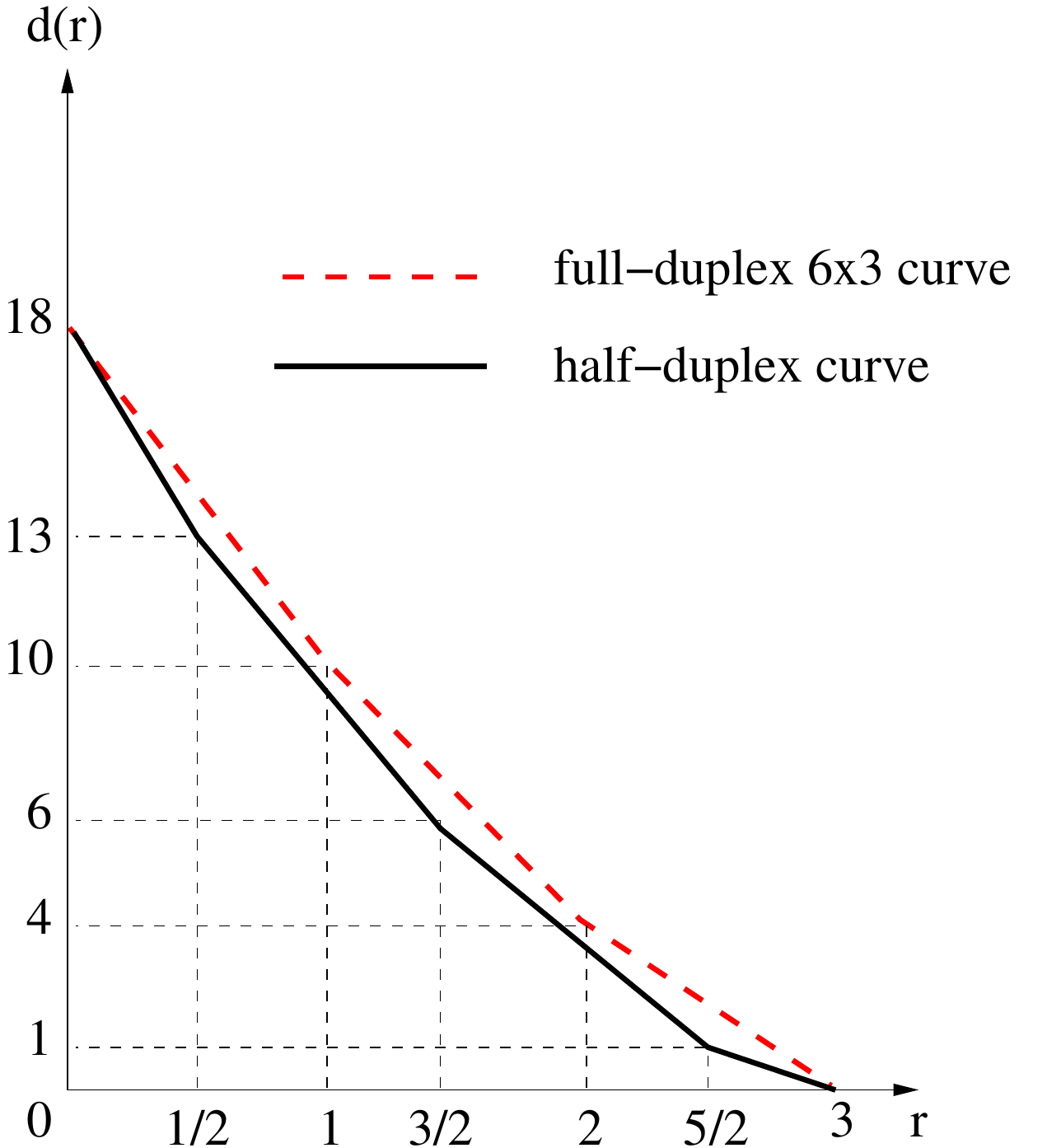} \hspace{1cm} \includegraphics[width=6cm]{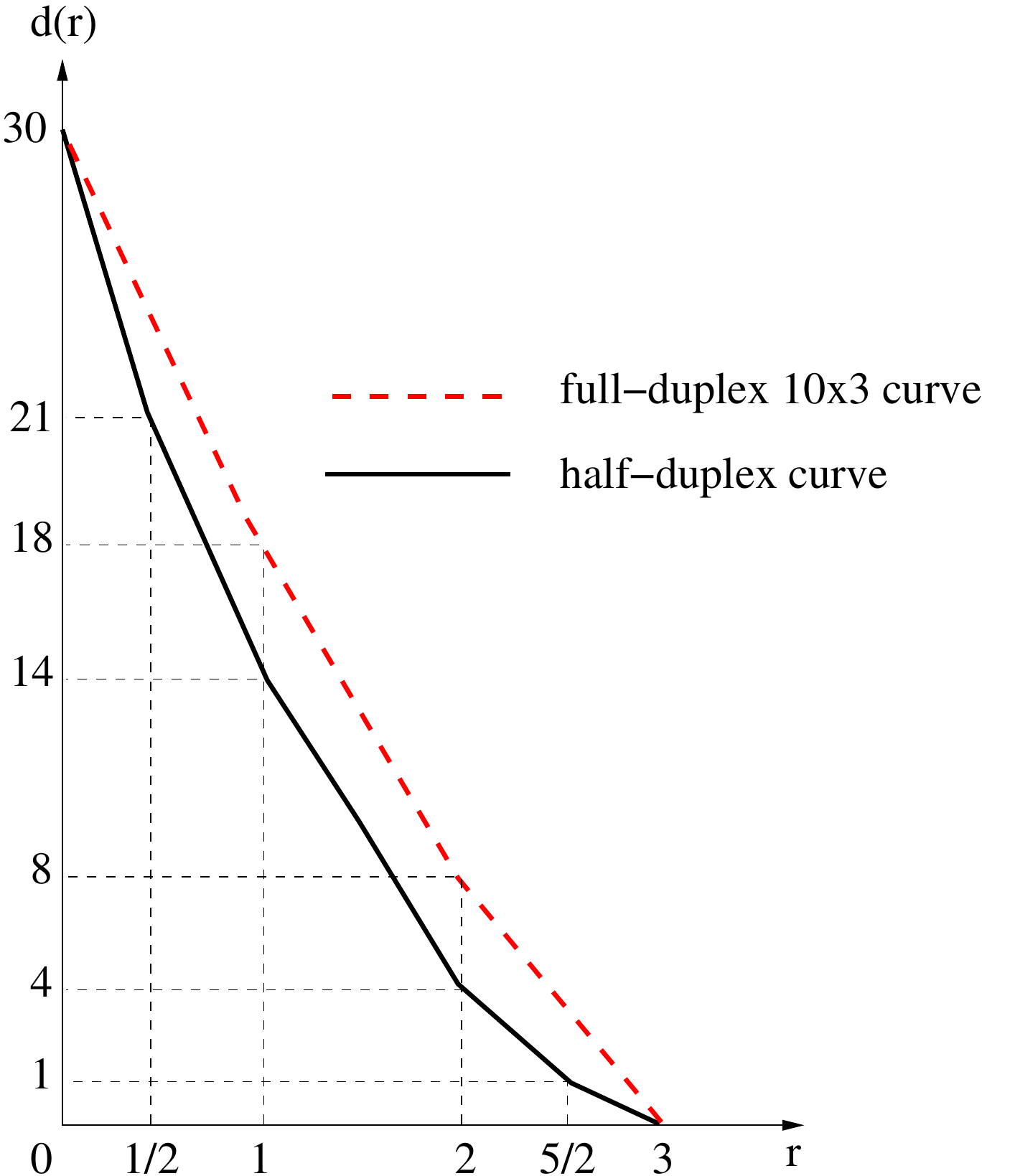}
\caption{Half-duplex DMT curve for $n=3$ and $m=1$ (left) / $n=3$ and $m=3$ (middle) / $n=3$ and $m=7$ (right).} \label{fig:dmt_curves}
\end{figure}

The results are illustrated in a particular case on Figure \ref{fig:dmt_curves}. DMT curves are shown for the case where the source and the destination have $n=3$ antennas each and the relay has $m=1$, $3$ and $7$ antennas, respectively. As expected, the loss due to the half-duplex constraint increases as $m$ gets larger.

A precise description of the main outage events at the corner points of the diversity curve is given below, along with the computation of the corresponding diversity. The main outage events are described in terms of the number of links being active in the channel matrices $G$ and $H$. For the direct channel matrix $G$, the number of active links is equal to the number of non-vanishing singular values $\sqrt{\lambda_j}$ (or rank) of $G$, that is, the number of $\alpha_j$ being equal to zero (since $\lambda_j=P^{-\alpha_j}$). For the relay channel matrix $H$, such a relation is less immediate, since the numbers $\beta_j$ are related to the singular values $\sqrt{\mu_j}$ of the {\em combined} channel matrix $(I_n+PGG^*)^{-1/2} H$ (through the equality $\mu_j=P^{-\beta_j}$). From this relation, we can actually deduce that the number of active links in the relay channel matrix $H$ is equal to the {\em sum} of the number of $\alpha_j$ and $\beta_j$ being equal to zero. In more detail, this gives rise to the following interpretation.

a) For $r=l/2$, where $l \in \{0,\ldots,l_0\}$, the main outage event is the event that only $l$ relay links are active, while all $n$ direct links are off. More precisely, we have
$$
\alpha_1^*=\ldots=\alpha_n^*=1 \quad \text{and} \quad \beta_1^*=\ldots=\beta_l^*=0, \, \beta_{l+1}^*=\ldots=\beta_{m \wedge n}^*=1
$$
These satisfy the constraints and the terms in (\ref{optprob}) become
\begin{align*}
& \sum_{j=1}^n (2n-2j+1) \, \alpha_j^* = n^2, \quad \sum_{j=l}^{m \wedge n} (n+m-2j+1) \, \beta_j^*
= (m-l)(n-l)\\
& m \sum_{j=1}^n (1-\alpha_j^*)^+ = 0,
\quad \sum_{j,k=1 \atop j+k \le n}^{n,m \wedge n} (1-\alpha_j^*-\beta_k^*)^+=0
\end{align*}
so the corresponding diversity is given by
$$ 
d_{\mathrm{HD},0}(l/2) = n^2 + (m-l)(n-l)
$$
b) For $r=l_0/2+l$, where $l \in \{0,\ldots,n-l_0\}$, the main outage event is that only $l$ direct links and $l_0+l$ relay links are active. More precisely, we have
$$
\alpha_1^*=\ldots=\alpha_l^*=0, \, \alpha_{l+1}^*=\ldots=\alpha_n^*=1 \quad \text{and} \quad \beta_1^*=\ldots=\beta_{l_0}^*=0, \, \beta_{l_0+1}^*=\ldots=\beta_{m \wedge n}^*=1
$$
These satisfy the constraints and the terms in (\ref{optprob}) become
\begin{align*}
& \sum_{j=1}^n (2n-2j+1) \, \alpha_j^* = (n-l)^2, \quad \sum_{j=l}^{m \wedge n} (n+m-2j+1) \, \beta_j^* = (m-l_0)(n-l_0)\\
& m \sum_{j=1}^n (1-\alpha_j^*)^+ = ml,
\quad \sum_{j,k=1 \atop j+k \le n}^{n,m \wedge n} (1-\alpha_j^*-\beta_k^*)^+ = l l_0
\end{align*}
so the corresponding diversity is given by
$$ 
d_{\mathrm{HD},0}(l_0/2+l) = (n+m-l)(n-l_0-l) + l_0^2
$$
c) For $r=n-l/2$, where $l \in \{0,\ldots,l_0/2\}$, the main outage event is that only $n-l$ direct links are active
and all $n$ relay links are active. More precisely, we have
$$
\alpha_1^*=\ldots=\alpha_{n-l}^*=0, \, \alpha_{n-l+1}^*=\ldots=\alpha_n^*=1 \quad \text{and} \quad \beta_1^*=\ldots=\beta_l^*=0, \, \beta_{l+1}^*=\ldots=\beta_{m \wedge n}^*=1
$$
These satisfy the constraints and the terms in (\ref{optprob}) become
\begin{align*}
& \sum_{j=1}^n (2n-2j+1) \, \alpha_j^* = l^2, \quad \sum_{j=1}^{m \wedge n} (n+m-2j+1) \, \beta_j^* = (m-l)(n-l)\\
& m \sum_{j=1}^n (1-\alpha_j^*)^+ = m(n-l),
\quad \sum_{j,k=1 \atop j+k \le n}^{n,m \wedge n} (1-\alpha_j^*-\beta_k^*)^+=(n-l)l
\end{align*}
so the corresponding diversity is given by
$$
d_{\mathrm{HD},0}(n-l/2) = l^2
$$
Let us end this section by a comment on the more general case, where the number of antennas at the source is not equal to the number of antennas at the destination. In this situation, it is clear that due to the inherent asymmetry of the problem, the optimal proportion $t$ where the relay is listening is not equal to $1/2$. In addition, this optimal proportion not only depends on $t$ but also on the target rate $r$. This makes the computation of the diversity order highly cumbersome in this case.


\section{New asymptotics of spherical integrals} \label{sec:spherical}
The goal of this section is to prove (\ref{asymptotic1}), namely that
$$
\lim_{P \to \infty} \frac{\log(\IC(P))}{\log P} = \left\{\begin{array}{l} -\infty, \quad \quad \text{if }
\exists j \le n \wedge p, \, k \le n \wedge m \text{ such that } j+k=n+1 \text{ and } \alpha_j+\beta_k<1\\ \\
\displaystyle - \sum_{j=1}^{n \wedge p} \sum_{k=1}^{n \wedge m} (1-\alpha_j-\beta_k)^+, \quad \text{otherwise}
\end{array} \right.
$$
where
$$
\IC(P) = \EE_U \left( \exp \left( -\sum_{j=1}^{n \wedge p} \sum_{k=1}^{n \wedge m} P^{1-\alpha_j-\beta_k} \, |u_{jk}|^2 \right) \right)
$$
Asymptotics of spherical integrals have been already studied in the mathematical literature (see for instance \cite{GZ01,GM05}) and have moreover found applications to the study of the performance of MMSE receivers in CDMA systems, when users employ different powers (this can be inferred from the paper \cite{TZ00}, although only the equal power case is considered in there). The asymptotic regime considered here is different from the one considered in the above references. In our case, the matrix sizes $n,m$ are fixed and the parameter $P$ tends to infinity, while in \cite{GZ01,GM05}, the matrix sizes $n,m$ tend to infinity. Accordingly, our approach departs substantially from the above mentioned papers, since it highly relies on the fact that the dimensions $n$ and $m$ are both fixed, while $P \to \infty$.

Notice that in the case where the matrix dimensions are fixed, an alternate computation technique for spherical integrals was already developed in \cite{YB07} in a slightly different context. In this work, the celebrated Harish-Chandra-Itzykson-Zuber formula was used in order to rewrite the integral as a determinant, and then the asymptotic analysis of the determinant was performed. This only works however when the dimensions $p$ and $m$ are greater than or equal to $n$.

$\bullet\;$Let us first show that if there exist $1 \le j \le n \wedge p$ and
$1 \le k \le n \wedge m$ such that $j+k=n+1$ and $\alpha_j+\beta_k<1$, then
\begin{equation} \label{suppol}
\EE_U \left( \exp \left( -\sum_{j=1}^{n \wedge p} \sum_{k=1}^{n \wedge m} P^{1-\alpha_j-\beta_k} \, |u_{jk}|^2 \right) \right) \le \exp(-P^c)
\end{equation}
where $c=\max\{1-\alpha_j - \beta_k \, : \, 1 \le j \le n \wedge p, \, 1 \le k \le n \wedge m, \, j+k=n+1\}>0$ (implying the first line of (\ref{asymptotic1})). Indeed,
\begin{eqnarray*}
\EE_U \left( \exp \left( -\sum_{j=1}^{n \wedge p} \sum_{k=1}^{n \wedge m} P^{1-\alpha_j-\beta_k} |u_{jk}|^2 \right) \right)
& \le & \exp \left(- \min_{U \in \UC(n)} \sum_{j=1}^{n \wedge p} \sum_{k=1}^{n \wedge m} P^{1-\alpha_j-\beta_k} |u_{jk}|^2 \right)\\
& \overset{(a)}{=} & \exp \left(- \sum_{j=1}^{n \wedge p} \sum_{k=1}^{n \wedge m} \delta_{j+k,n+1} \, P^{1-\alpha_j-\beta_k} \right) \le \exp(-P^c)
\end{eqnarray*}
where $c>0$ is defined as above. The equality (a) follows from the fact that
if $\lambda_1 \ge \ldots \ge \lambda_{n \wedge p}$ and $\mu_1 \ge \ldots \ge \mu_{n \wedge m}$, then $\{u_{jk}=\delta_{j+k,n+1}\}$ minimizes
$$
\sum_{j=1}^{n \wedge p} \sum_{k=1}^{n \wedge m} \lambda_j \mu_k |u_{jk}|^2
$$
over all the matrices in $\UC(n)$. This shows (\ref{suppol}).


$\bullet\;$Since we know that the integral decays super-polynomially with $P$ if  $c>0$, let us now assume that
\begin{equation} \label{temp_assump}
\alpha_j+\beta_k \ge 1, \quad \forall 1 \le j \le n \wedge p, \, 1 \le k \le n \wedge m \text{ such that } j+k=n+1
\end{equation}
Because of the ordering of the $\alpha$'s and $\beta$'s, this implies that $\alpha_j + \beta_k \ge 1$ for all $j,k$ such that $j+k \ge n+1$. Let us also define
$$
X(P) = \sum_{j=1}^{n \wedge p} \sum_{k=1 \atop \hspace{-1cm} \{j+k \le n, \, \alpha_j + \beta_k < 1\}}^{n \wedge m} \, P^{1-\alpha_j-\beta_k} |u_{jk}|^2
$$
It can first be checked that under the assumption (\ref{temp_assump}),
$$
\IC(P) \doteq \EE(\exp(-X(P)))
$$
Indeed, it holds under the same assumption that
$$
X(P) \le \sum_{j=1}^{n \wedge p} \sum_{k=1}^{n \wedge m} P^{1-\alpha_j-\beta_k} |u_{jk}|^2 \le X(P) + n^2
$$
so
$$
\EE(\exp(-X(P))) \ge \IC(P) \ge \EE(\exp(-X(P))) \, e^{-n^2}
$$
$\bullet\;$For each $1 \le k \le (n-1) \wedge m$, let us now define $j_k=\sup\{1 \le j \le (n-k) \wedge p \, : \, \alpha_j + \beta_k < 1\}$. Notice that
$$
\alpha_j+\beta_k < 1, \quad \forall 1 \le j \le j_k
$$
because of the ordering of the $\alpha$'s, and also that
$$
j_k \le j_l, \quad \text{if } l \le k
$$
because of the ordering of the $\beta$'s. Moreover, $X(P)$ may be rewritten as
$$
X(P) = \sum_{k=1}^{(n-1) \wedge m} \sum_{j=1}^{j_k} P^{1-\alpha_j-\beta_k} |u_{jk}|^2
$$
$\bullet\;$Next, we compute
$$
\EE(\exp(-X(P))) = \int_0^1 \PP( \exp(-X(P)) \ge t) \, dt = \int_0^1 \PP(X(P) \le \log(1/t)) \, dt
$$
From this, we deduce that
$$
\EE(\exp(-X(P))) \ge \int_0^{\frac{1}{e}} \PP(X(P) \le \log(1/t)) \, dt \ge \frac{1}{e} \, \PP(X(P) \le 1) 
$$
and that for any $0<\varepsilon<1$,
\begin{equation} \label{proof_temp1}
\EE(\exp(-X(P))) \le \varepsilon + \int_\varepsilon^1 \PP(X(P) \le \log(1/t)) \, dt \le \varepsilon + \PP( X(P) \le \log(1/\varepsilon))
\end{equation}
Remember that our aim is to show that $\EE(\exp(-X(P))) \doteq P^{-d}$, where
$$
d = \sum_{j=1}^{n \wedge p} \sum_{k=1}^{n \wedge m} (1-\alpha_j-\beta_k)^+ = \sum_{k=1}^{(n-1) \wedge m} \sum_{j=1}^{j_k} (1-\alpha_j-\beta_k)
$$
It is therefore natural to choose $\varepsilon=P^{-d}$ in (\ref{proof_temp1}), which leads to
$$
\frac{1}{e} \, \PP(X(P) \le 1) \le \EE(\exp(-X(P))) \le P^{-d} + \PP ( X(P) \le d \log P)
$$
If we then show that
\begin{equation} \label{proof_temp2}
\PP(X(P) \le 1) \doteq \PP ( X(P) \le d \log P) \doteq P^{-d}
\end{equation}
we will have proved (\ref{asymptotic1}).

$\bullet\;$At this stage, it is worth giving an intuition as to why (\ref{proof_temp2}) should hold, focusing on the case where $m, p \ge n$ for simplicity. It can first be checked that
$$
\PP(X(P) \le 1) \doteq \PP \left( |u_{jk}|^2 \le P^{-(1-\alpha_j-\beta_k)}, \; \forall 1 \le j \le j_k, \; 1 \le k \le n-1 \right)
$$
Being part of an $n \times n$ unitary matrix $U$, the $n^2$ random variables $|u_{jk}|^2$ are linked through $n(n+1)/2$ constraints. Nevertheless, there are no more than $n(n-1)/2$ such random variables involved in the above expression, so the joint probability that they are all asymptotically small is not affected by these constraints, and we may as well consider them as independent in this limit. This leads us to the following estimate:
$$
\PP(X(P) \le 1) \doteq \prod_{k=1}^{n-1} \prod_{j=1}^{j_k} \PP\left( |u_{jk}|^2 \le P^{-(1-\alpha_j-\beta_k)}\right) \doteq \prod_{k=1}^{n-1} \prod_{j=1}^{j_k}P^{-(1-\alpha_j-\beta_k)}
$$
which is the desired result (in the case where $m, p \ge n$). The rigorous proof follows.

$\bullet\;$Notice that since 
$$
\PP(X(P) \le 1) \le \PP \left( \sum_{j=1}^{j_k} P^{1-\alpha_j-\beta_k} |u_{jk}|^2 \le 1, \; \forall 1 \le k \le (n-1) \wedge m \right) \le \PP(X(P) \le n)
$$
it holds that
$$
\PP(X(P) \le 1) \doteq \PP \left( \sum_{j=1}^{j_k} P^{1-\alpha_j-\beta_k} |u_{jk}|^2 \le 1, \; \forall 1 \le k \le (n-1) \wedge m \right)
$$
and similarly that
$$
\PP( X(P) \le d \log P) \doteq \PP \left( \sum_{j=1}^{j_k} P^{1-\alpha_j-\beta_k} |u_{jk}|^2 \le d \log P, \; \forall 1 \le k \le (n-1) \wedge m \right)
$$
Proving (\ref{proof_temp2}) therefore boils down to proving that
\begin{equation} \label{asymptotic2}
\PP \left( \sum_{j=1}^{j_k} P^{1-\alpha_j-\beta_k} |u_{jk}|^2 \le s, \; \forall 1 \le k \le (n-1) \wedge m \right) \doteq P^{-d}
\end{equation}
for $s=$ either $1$ or $d \log P$. This asymptotic equality may be recast into the following form:
$$
\PP \left( \sum_{j=1}^{j_k} \frac{|u_{jk}|^2}{s_{jk}} \le 1, \; \forall 1 \le k \le (n-1) \wedge m \right) \doteq \prod_{k=1}^{(n-1) \wedge m} \prod_{j=1}^{j_k} s_{jk}
$$
where $s_{jk}=$ either $1/P^{1-\alpha_j-\beta_k}$ or $(d \log P)/P^{1-\alpha_j-\beta_k}$ is tending to zero as $P \to \infty$, if $j \le j_k$ (notice that when present, the $\log P$ term in the $s_{jk}$'s on the right-hand side does not affect the asymptotic polynomial decay in $P$).

$\bullet\;$In order to estimate the above probability, let us expand it as
\begin{eqnarray*}
\lefteqn{\hspace{-2.3cm}\PP \left( \sum_{j=1}^{j_k} \frac{|u_{jk}|^2}{s_{jk}} \le 1, \; \forall 1 \le k \le (n-1) \wedge m \right)
= \PP \left( \sum_{j=1}^{j_1} \frac{|u_{j1}|^2}{s_{j1}} \le 1 \right) \times \PP \left( \sum_{j=1}^{j_2} \frac{|u_{j2}|^2}{s_{j2}} \le 1 \; \Bigg| \; 
\sum_{j=1}^{j_1} \frac{|u_{j1}|^2}{s_{j1}} \le 1 \right)}\\
& \times & \ldots\\
& \times & \PP \left( \sum_{j=1}^{j_k} \frac{|u_{jk}|^2}{s_{jk}} \le 1 \; \Bigg| \; 
\sum_{j=1}^{j_l} \frac{|u_{jl}|^2}{s_{jl}} \le 1, \; \forall 1 \le l < k \right)\\
& \times & \ldots\\
& \times & \PP \left( \sum_{j=1}^{j_{(n-1) \wedge m}} \frac{|u_{j,(n-1) \wedge m}|^2}{s_{j,(n-1) \wedge m}} \le 1 \; \Bigg| \; 
\sum_{j=1}^{j_l} \frac{|u_{jl}|^2}{s_{jl}} \le 1, \; \forall 1 \le l < (n-1) \wedge m \right)
\end{eqnarray*}

If we therefore prove that for all $1 \le k \le (n-1) \wedge m$,
\begin{equation} \label{asymptotic3}
\PP \left( \sum_{j=1}^{j_k} \frac{|u_{jk}|^2}{s_{jk}} \le 1 \; \Bigg| \; 
\sum_{j=1}^{j_l} \frac{|u_{jl}|^2}{s_{jl}} \le 1, \; \forall 1 \le l < k \right) \doteq \prod_{j=1}^{j_k} s_{jk}
\end{equation}
we will have proved (\ref{asymptotic2}). We will show the above inequality step by step, starting with the simplest case $k=1$.

$\bullet\;$The rest of the proof relies on the following fact (see for instance Lemma 2.2. in \cite{PetzReffy04}): a matrix $U$ which is Haar distributed on $\UC(n)$ may be obtained from the Gram-Schmidt orthonormalization of an $n \times n$ matrix $G$ with i.i.d.~$\NC_\CC(0,1)$ entries, that is:
$$
\begin{array}{ll}
u_1 = w_1/\| w_1 \|, & \text{where } w_1 = g_1\\
& \\
u_2 = w_2/\|w_2\|, & \text{where } w_2 = g_2 - (u_1^* g_2) \, u_1\\
& \\
u_k=w_k/\| w_k \|, & \text{where } w_k = g_k - \sum_{l=1}^{k-1} (u_l^* g_k) \, u_l\\
& \\
\end{array}
$$
$u_k,w_k,g_k$ are the column vectors of the matrices $U,W,G$, respectively. A key observation is that conditioned on the column vectors $u_1,...,u_{k-1}$, the column vector $w_k$ is a Gaussian vector with zero mean and covariance matrix $I_n-\sum_{l=1}^{k-1} u_l u_l^*$.


$\bullet\;${\it Step 1. First column (k=1).} We need to compute
$$
\PP \left( \sum_{j=1}^{j_1} \frac{|u_{j1}|^2}{s_{j1}} \le 1 \right) =
\PP \left( \sum_{j=1}^{j_1} \frac{|w_{j1}|^2}{s_{j1}} \le \| w_1 \|^2 \right)\\
= \PP \left( \sum_{j=1}^{j_1} \left( \frac{1-s_{j1}}{s_{j1}} \right) \, |w_{j1}|^2 \le \sum_{j=j_1+1}^n |w_{j1}|^2 \right)
$$
where $w_{11}, \ldots, w_{n1}$ are i.i.d.$\sim\NC_\CC(0,1)$ random variables.
For this, we need the following lemma, whose proof is relegated to the appendix.

\begin{lem} \label{lemtemp}
Let $A$ be a positive definite $m \times m$ matrix, $\lambda_{\min}(A)$ denote its smallest eigenvalue and $y \sim \NC_\CC(0,I_m)$, $z \sim \NC_\CC(0,I_n)$ be independent Gaussian vectors. Then for all $\lambda_0>0$, there exist $C_2>C_1>0$ independent of the matrix $A$ (but possibly dependent on the parameters $m$ and $n$) such that if $\lambda_{\min}(A) \ge \lambda_0$, then 
$$
\frac{C_1}{\det A} \le \PP (y^*Ay \le \|z\|^2) \le \frac{C_2}{\det A}.
$$
\end{lem}

Remembering that the $s_{j1}$ tend to zero as $P \to \infty$, we can therefore apply the above lemma with $A=\text{diag} \left( \frac{1-s_{11}}{s_{11}},\ldots,\frac{1-s_{j_1,1}}{s_{j_1,1}} \right)$ to obtain
$$
\PP \left( \sum_{j=1}^{j_1} \left( \frac{1-s_{j1}}{s_{j1}} \right) \, |w_{j1}|^2 \le \sum_{j=j_1+1}^n |w_{j1}|^2 \right)
\doteq \prod_{j=1}^{j_1} \frac{s_{j1}}{1-s_{j1}} \doteq \prod_{j=1}^{j_1} s_{j1}.
$$
which is the desired result for $k=1$.


$\bullet\;${\it Step 2. Second column (k=2).} We need to compute
$$
\PP \left( \sum_{j=1}^{j_2} \frac{|u_{j2}|^2}{s_{j2}} \le 1 \; \Bigg| \; 
\sum_{j=1}^{j_1} \frac{|u_{j1}|^2}{s_{j1}} \le 1 \right)
= \PP \left( \PP \left( \sum_{j=1}^{j_2} \frac{|w_{j2}|^2}{s_{j2}} \le \| w_2 \|^2 \; \Bigg| \; u_1 \right) \; \Bigg| \; 
\sum_{j=1}^{j_1} \frac{|u_{j1}|^2}{s_{j1}} \le 1 \right)
$$
Remember that conditioned on $u_1$, $w_2 \sim \NC_\CC(0,I_n-u_1 u_1^*)$. Since rank$(I_n-u_1u_1^*)=n-1$, this covariance matrix may be rewritten as $I_n-u_1u_1^*=VV^*$, where $V$ is an $n \times (n-1)$ matrix; notice that the $n \times n$ matrix $(u_1|V)$ is unitary, since $u_1 u_1^* + VV^*=I_n$, and that $V^*V=I_{n-1}$.

The Gaussian vector $w_2$ may also be rewritten as $w_2 = V x_2$, where $x_2 \sim \NC_\CC(0,I_{n-1})$. Noticing that $\| w_2 \|^2 = x_2^* V^* V x_2 = \| x_2 \|^2$, we obtain
$$
\PP \left( \sum_{j=1}^{j_2} \frac{|w_{j2}|^2}{s_{j2}} \le \| w_2 \|^2 \; \Bigg| \; u_1 \right)
= \PP \left( \sum_{j=1}^{j_2} \frac{|(Vx_2)_j|^2}{s_{j2}} \le \| x_2 \|^2 \right)
$$
since the distribution of $x_2$ does not depend on $u_1$. Let now $D=\text{diag}(s_{12}^{-1},\ldots,s_{j_2,2}^{-1})$, $\tilde{V}$ be the $j_2 \times (n-1)$ matrix composed of the first $j_2$ rows of $V$ and $\tilde{u}_1$ be the vector $u_1$ restricted to its first $j_2$ components. We see that $\tilde{V} \tilde{V}^*=I_{j_2}-\tilde{u}_1 \tilde{u}_1^*$ and that
$$
\PP \left( \sum_{j=1}^{j_2} \frac{|(Vx_2)_j|^2}{s_{j2}} \le \| x_2 \|^2 \right)
= \PP \left( x_2^* \tilde{V}^* D \tilde{V} x_2 \le \| x_2 \|^2 \right)
$$
Using the $LQ$ decomposition of the matrix $\tilde{V}$ (i.e. $\tilde{V}=LQ$, where $L$ is a lower-triangular $j_2 \times (n-1)$ matrix and $Q$ is a unitary $(n-1) \times (n-1)$ matrix) together with the fact that the distribution of $x_2$ is unitarily invariant, we obtain
$$
\PP \left( x_2^* \tilde{V}^* D \tilde{V} x_2 \le \| x_2 \|^2 \right) 
= \PP \left( x_2^* L^* D L x_2 \le \| x_2 \|^2 \right)
$$
Writing now $L=(\tilde{L}|0)$, where $\tilde{L}$ is a square $j_2 \times j_2$ matrix, and $x_2=(y_2,z_2)$, where $y_2$ are the first $j_2$ components of $x_2$, and $z_2$ are the last $n-1-j_2$ ones (remember that $n-1-j_2 \ge 1$), we see that the above probability may be rewritten as
$$
\PP \left( x_2^* L^* D L x_2 \le \| x_2 \|^2 \right)
= \PP \left( y_2^* \tilde{L}^* D \tilde{L} y_2 \le \| y_2 \|^2 + \| z_2 \|^2 \right)
= \PP \left( y_2^* (\tilde{L}^* D \tilde{L} - I_{j_2}) y_2 \le z_2^* z_2 \right)
$$
Using again Lemma \ref{lemtemp}, we deduce that
$$
\PP \left( y_2^* \left( \tilde{L}^* D \tilde{L} - I_{j_2} \right) y_2 \le z_2^* z_2 \right) \doteq
\frac{1}{\det(\tilde{L}^* D \tilde{L}-I_{j_2})} \doteq \frac{1}{\det(\tilde{L}^* D \tilde{L})}
= \frac{1}{\det(D) \, \det(\tilde{L}\tilde{L}^*)} \doteq \prod_{j=1}^{j_2} s_{j2}
$$
Indeed,
$$
\det(\tilde{L} \tilde{L}^*) = \det(LL^*) = \det(\tilde{V}\tilde{V}^*) \in [1 - \|\tilde{u}_1\|^2,1]
$$
and the condition $\sum_{j=1}^{j_1} \frac{|u_{j1}|^2}{s_{j1}} \le 1$ implies that $|u_{j1}|^2 \le s_{j1}$ for all $1 \le j \le j_1$, in particular for all $1 \le j \le j_2$, and therefore that $1-\| \tilde{u}_1 \|^2$ is asymptotically close to $1$. This concludes the proof for $k=2$.


$\bullet\;${\it Step 3. k-th column.} We repeat the procedure of Step 2 in this more general setting. We need to compute
\begin{eqnarray*}
\lefteqn{\PP \left( \sum_{j=1}^{j_k} \frac{|u_{jk}|^2}{s_{jk}} \le 1 \; \Bigg| \; 
\sum_{j=1}^{j_l} \frac{|u_{jl}|^2}{s_{jl}} \le 1, \; \forall 1 \le l < k \right)}\\
& = &  \quad \PP \left( \PP \left( \sum_{j=1}^{j_k} \frac{|w_{jk}|^2}{s_{jk}} \le \| w_k \|^2 \; \Bigg| \; u_1,\ldots,u_{k-1} \right) \; \Bigg| \; 
\sum_{j=1}^{j_l} \frac{|u_{jl}|^2}{s_{jl}} \le 1, \; \forall 1 \le l < k  \right)
\end{eqnarray*}
Remember that conditioned on $u_1,\ldots,u_{k-1}$, $w_k \sim \NC_\CC(0,I_n-\sum_{l=1}^{k-1} u_l u_l^*)$. Since rank$(I_n-\sum_{l=1}^{k-1}u_lu_l^*)=n-k+1$, this covariance matrix may be rewritten as $I_n-\sum_{l=1}^{k-1}u_lu_l^*=VV^*$, where $V$ is an $n \times (n-k+1)$ matrix; notice that the $n \times n$ matrix $(u_1|\ldots|u_{k-1}|V)$ is unitary, since $\sum_{l=1}^{k-1} u_l u_l^* + VV^*=I_n$, and that $V^*V=I_{n-k+1}$.

The Gaussian vector $w_k$ may also be rewritten as $w_k = V x_k$, where $x_k \sim \NC_\CC(0,I_{n-k+1})$. Noticing that $\| w_k \|^2 = x_k^* V^* V x_k = \| x_k \|^2$, we obtain
$$
\PP \left( \sum_{j=1}^{j_k} \frac{|w_{jk}|^2}{s_{jk}} \le \| w_k \|^2 \; \Bigg| \; u_1, \ldots, u_{k-1} \right)
= \PP \left( \sum_{j=1}^{j_k} \frac{|(Vx_k)_j|^2}{s_{jk}} \le \| x_k \|^2 \right)
$$
since the distribution of $x_k$ does not depend on $u_1, \ldots, u_{k-1}$. Let now $D=\text{diag}(s_{1k}^{-1},\ldots,s_{j_k,k}^{-1})$, $\tilde{V}$ be the $j_k \times (n-1)$ matrix composed of the first $j_k$ rows of $V$ and $\tilde{u}_1,\ldots,\tilde{u}_{k-1}$ be the vectors $u_1,\ldots,u_{k-1}$ restricted to their first $j_k$ components. We see that $\tilde{V} \tilde{V}^*=I_{j_k}-\sum_{l=1}^{k-1} \tilde{u}_l \tilde{u}_l^*$ and that
$$
\PP \left( \sum_{j=1}^{j_k} \frac{|(Vx_k)_j|^2}{s_{jk}} \le \| x_k \|^2 \right)
= \PP \left( x_k^* \tilde{V}^* D \tilde{V} x_k \le \| x_k \|^2 \right)
$$
Using the $LQ$ decomposition of the matrix $\tilde{V}$ (i.e. $\tilde{V}=LQ$, where $L$ is a lower-triangular $j_k \times (n-1)$ matrix and $Q$ is a unitary $(n-1) \times (n-1)$ matrix) together with the fact that the distribution of $x_k$ is unitarily invariant, we obtain
$$
\PP \left( x_k^* \tilde{V}^* D \tilde{V} x_k \le \| x_k \|^2 \right) 
= \PP \left( x_k^* L^* D L x_k \le \| x_k \|^2 \right)
$$
Writing now $L=(\tilde{L}|0)$, where $\tilde{L}$ is a square $j_k \times j_k$ matrix, and $x_k=(y_k,z_k)$, where $y_k$ are the first $j_k$ components of $x_k$, and $z_k$ are the last $n-k+1-j_k$ ones (remember that $n-k+1-j_k \ge 1$), we see that the above probability may be rewritten as
$$
\PP \left( x_k^* L^* D L x_k \le \| x_k \|^2 \right)
= \PP \left( y_k^* \tilde{L}^* D \tilde{L} y_k \le \| y_k \|^2 + \| z_k \|^2 \right)
= \PP \left( y_k^* (\tilde{L}^* D \tilde{L} - I_{j_k}) y_k \le z_k^* z_k \right)
$$
Using again Lemma \ref{lemtemp}, we deduce that
$$
\PP \left( y_k^* \left( \tilde{L}^* D \tilde{L} - I_{j_k} \right) y_k \le z_k^* z_k \right) \doteq
\frac{1}{\det(\tilde{L}^* D \tilde{L}-I_{j_k})} \doteq \frac{1}{\det(\tilde{L}^* D \tilde{L})}
= \frac{1}{\det(D) \, \det(\tilde{L}\tilde{L}^*)} \doteq \prod_{j=1}^{j_2} s_{j2}
$$
Indeed,
$$
\det(\tilde{L} \tilde{L}^*) = \det(LL^*) = \det(\tilde{V}\tilde{V}^*) \in
\left[1 - \sum_{l=1}^{k-1} \|\tilde{u}_l\|^2,1 \right]
$$
and the condition $\sum_{j=1}^{j_l} \frac{|u_{jl}|^2}{s_{jl}} \le 1$ for all $1 \le l <k$ implies that $|u_{jl}|^2 \le s_{jl}$ for all $1 \le l \le k$, $1 \le j \le j_l$, in particular for all $1 \le j \le j_k$, and therefore that $1- \sum_{l=1}^{k-1} \| \tilde{u}_l \|^2$ is asymptotically close to $1$. This proves (\ref{asymptotic3})
and therefore concludes the proof of (\ref{asymptotic1}).
\hfill $\blacksquare$


\section{Conclusion and Perspectives}\label{sec:Conclusion}
In this work, we have expressed the half-duplex relay channel diversity-multiplexing tradeoff (DMT) as the solution of a minimization problem.
In the symmetric case, i.e.~when the source and the destination have $n$ antennas each and the relay has $m$ antennas, the solution to the minimization problem is given explicitly. Static relaying protocols are considered. Our results show that the static half-duplex DMT is equal to the full-duplex DMT only when the relay has a single antenna. In all other cases, performance losses are to be expected because of the half-duplex constraint. The case of the dynamic half-duplex relay is more complex and left open.

We believe that the asymptotics of spherical integrals developed in this paper for the study of the half-duplex DMT are of interest in their own right, as such asymptotics were not considered before in the literature. Besides, as mentioned
at the beginning of Section \ref{sec:method}, the problem of determining the joint eigenvalue distribution of two correlated Wishart matrices remains open.

\section*{Acknowledgment}
We are very grateful to Paul Cuff and Elza Erkip who contributed at the beginning of this work on the topic and came up with interesting comments and questions along the development of the paper. In addition, we would like to thank both Emre Telatar and Ofer Zeitouni for insightful comments that were of great help for the progress of this work.


\appendix

{\em Proof of Lemma \ref{lemtemp}.}
First notice that
$$
\PP( y^* A y \le \|z\|^2) = \int_0^\infty \PP(y^*Ay \le t) \, f_{\|z\|^2}(t) \, dt
$$
where $f_{\|z\|^2}(t) = t^{n-1} \, \exp(-t) / (n-1)!$ and
$$
\PP( y^* A y \le t) = \int_{\{y \in \CC^m \, : \, y^* A y \le t\}} \frac{1}{\pi^m} \,
\exp(- \|y \|^2) \, dy = \frac{1}{\pi^m \, \det A} \int_{\{\xi \in \CC^m \, : \, \| \xi \|^2 \le t\}} \exp(-\xi^* A^{-1} \xi)
\, d\xi
$$
by the change of variable $\xi = A^{1/2} y$. This probability is further upper bounded by
$$
\PP( y^* A y \le t) \le \frac{1}{\pi^m \, \det A} \, \left| \left\{\xi \in \CC^m: \|\xi\|^2 \le t \right\} \right|
= \frac{C_m \, t^m}{\pi^m \, \det A}
$$
where $C_m = \left| \left\{\xi \in \CC^m: \|\xi\|^2 \le 1 \right\} \right|$. It is also lower bounded by
\begin{eqnarray*}
\PP( y^* A y \le t) & \ge & \frac{1}{\pi^m \, \det A} \int_{\{\xi \in \CC^m \, : \, \| \xi \|^2 \le t\}} \exp \left( -\frac{\|\xi\|^2}{\lambda_{\min}(A)} \right) \, d\xi\\
& \ge & \frac{1}{\pi^m \, \det A} \, \left| \left\{\xi \in \CC^m: \|\xi\|^2 \le t \right\} \right|\, \exp\left( - \frac{t}{\lambda_{\min}(A)} \right) = \frac{C_m \, t^m}{\pi^m \, \det A} \, \exp \left( - \frac{t}{\lambda_{\min}(A)} \right)
\end{eqnarray*}
In total, we therefore obtain
$$
\PP( y^* A y \le \|z\|^2) \le \frac{C_m}{\pi^m \, \det A} \int_0^\infty \frac{t^{m+n-1} \, \exp(-t)}{(n-1)!} \, dt
= \frac{C_m \, (m+n-1)!}{\pi^m \, (n-1)!} \, \frac{1}{\det(A)}
$$
and
\begin{eqnarray*}
\PP( y^* A y \le \|z\|^2) & \ge & \frac{C_m}{\pi^m \, \det A} \int_0^\infty \frac{t^{m+n-1} \, \exp\left(-\left(1+\frac{1}{\lambda_{\min}(A)}\right)t\right)}{(n-1)!} \, dt\\
& = & \frac{C_m}{\pi^m \, \det A} \, \left(\frac{\lambda_{\min}(A)}{1+\lambda_{\min}(A)}\right)^{m+n} \int_0^\infty \frac{s^{m+n-1} \, \exp(-s)}{(n-1)!} \, ds\\
& = & \frac{C_m \, (m+n-1)!}{\pi^m \, (n-1)!} \, \left( \frac{\lambda_{\min}(A)}{1+\lambda_{\min}(A)} \right)^{m+n} \,  \frac{1}{\det(A)}
\end{eqnarray*}
where we have used the change of variable $s= \left(\frac{\lambda_{\min}(A)+1}{\lambda_{\min}(A)}\right) t$. This proves the lemma, since by assumption, $\lambda_{\min}(A) \ge \lambda_0>0$.
\hfill $\blacksquare$


\end{document}